\documentclass[aps,draft,amsmath,amssymb]{revtex4} %FOR GRG STYLE
\def\be{\begin{equation}}
\def\ee{\end{equation}}
\def\beq{\begin{eqnarray}}
\def\eeq{\end{eqnarray}}

\def\bay{\begin{array}}
\def\eay{\end{array}}

\begin{document}

\preprint{smw-comment-01-06}
\title{On the significance of the Braxmaier et al.\ Laser
Interferometry \\ Experiment of 2002 for a Theory of Relativity}

\author{Sanjay M. Wagh}
\affiliation{Central India Research Institute, Post Box 606,
Laxminagar, Nagpur 440 022, India \\ E-mail:
cirinag\_ngp@sancharnet.in}

\date{June 2, 2006}
\begin{abstract}
A recent laser interferometry experiment %[Phys. Rev. Lett., Vol.
%88, 10401 (2002)]
\cite{braxmaier} registered a non-detection of a frequency shift
that amounts to a fractional change in the speed of light $\delta
c/c \approx (4.8\pm5.3)\times10^{-12}$. We comment here on the
theoretical significance of this experimental result.
%\centerline{Submitted to: General Relativity \& Gravitation}
\end{abstract}
\maketitle

An innovative laser interferometry experiment, wherein the
frequencies of two sources were compared over a period of $\sim$
200 days, used \cite{braxmaier} sources of the following
characteristics.

An Nd:YAG laser (frequency stabilized by cryogenically cooled,
ultra-stable, Nd:YAG laser using the Pound-Drever-Hall method
\cite{lase}) served as the ``length standard'' for the
interference experiment since the standing wave-mode number, $n$,
of the resonance cavity was held fixed. In this case, the size $L$
of the cavity controls the wavelength of the emitted light as:
$\lambda =2L/n$.

Another Nd:YAG laser (frequency stabilized to a normally forbidden
transition of gaseous iodine) served as the ``time standard'' for
the experiment. In this case, the original beam of emission was
split into two by a polarizing beam-splitter. The stronger of the
parts was then used as the pump beam and the weaker part as a
probe beam. The probe beam attenuation by the Iodine gas actively
controlled the resonator and stabilized it on resonance frequency
with the used Iodine transition.

Thus, if the length-standard's frequency varied independently of
that of the time-standard, then the relative frequency shift of
these sources would result. So, the $\lambda \approx 1\,\mu{\rm
m}$ emission of two sources, one as the length-standard and
another as an ``independent'' time-standard, was allowed
\cite{braxmaier} to form the interference to measure their
frequency difference via the fringe shifts.

This interesting  laser interferometry experiment then showed
\cite{braxmaier} a non-detection of a frequency shift, to be
specific, $\delta\nu/\nu \approx (4.8\pm5.3)\times10^{-12}$ over
an observation period of $\sim 200$ days. This is then also the
fractional change, $\delta c/c$, in the speed of light, $c$,
indicating that this {\em constant\/} of special relativity is
unaffected by factors like the motion of the Earth relative to the
reference system of the cosmic microwave background radiation
\cite{braxmaier, lieu-greg}.

In \cite{lieu-greg}, this result has been shown \footnote{A
measured value was $\delta\nu \sim 1.36\pm 0.63$ kHz, while that
implied by the (flat) standard model of cosmology is
$\delta\nu\sim6$ kHz. One could, however, argue that the
cosmological expansion has already ``decoupled'' from the locally
measurable effects.} to mean no or very little expansion of the
space, quite contrary to the well-known interpretation of the
galactic redshifts and an implication of the standard cosmology
\cite{cosmo} based on Einstein's field equations of general
relativity. It has thus been used \cite{lieu-greg} to question the
reality of space expansion at the rate inferred from the WMAP data
\cite{spergel}.

Of specific interest to us here is the absence of any noticeable
effects of the non-inertial character of the Earth's reference
system on the frequency shift. Not only is the Earth in motion
relative to the cosmic microwave background, it is rotating about
an axis as well. The absence of any effects of these motions on
the frequency-shift should be of fundamental significance.

Recall that, along with other experiments, the null result of the
Michelson-Morley experiment indicated \cite{lemwd, ein-pop} the
spurious nature of the ``ether.'' In the same spirit, the
non-detection of the frequency-shift in the experiment of
\cite{braxmaier} indicates that the speed of light (in vacuum) is
a {\em fundamental constant\/} that is independent of whether the
system of reference is inertial or non-inertial.

As was shown by Einstein \cite{lemwd}, the constancy of the speed
of light (in vacuum) (is inconsistent with the Galilean
transformations of Newton's theory but it) {\em can be made
consistent\/} with the special principle of relativity using the
Lorentz transformations, the special principle also being the
basis of the Galilean transformations. Similarly, if this
constancy is the above fundamental law, as it seems from
\cite{braxmaier}, then it needs to be made consistent with the
general principle of relativity by selecting a proper mathematical
formulation for this principle \footnote{Point transformations of
coordinates could be used to implement the constancy of the speed
of light. However, such a formalism will not be satisfactory in
view of Descartes's concern discussed later.}. It is but easily
seen that Einstein's field equations of the general relativity are
not consistent with this {\it complete\/} constancy of the speed
of light.

Einstein, likewise with Newton, had aimed at a comprehensive
theoretical description of all the physical phenomena. Thus, his
field equations of the general relativity {\em describe the
physical entirety \/} on representing ``gravity'' by the Einstein
tensor and on incorporating the ``remaining'' forces by way of the
energy-momentum tensor. Of course, only ``gravity'' gets
represented by the spacetime curvature in these, now famous,
Einstein's field equations.

But, to go beyond Newton's theory, the concept of force needs
(logical) replacement, and not any specific type of force like
gravity. It then follows that Einstein's equations are \cite{ppt}
logically unacceptable. The solutions of Einstein's field
equations are then inappropriate for describing any physical
situations. That is to say, various (non-Newtonian) properties of
the solutions of Einstein's field equations should be irrelevant
or inconsequential to any of the physical situations.

This above conclusion is inescapable even when Einstein's field
equations provide, as solutions, 4-dimensional (spacetime)
geometries, mathematical constructions, which only are
mathematically well-defined but physically certainly dubious.

In fact, by 1928, Einstein had concluded \cite{schlipp, pais} that
his field equations were not any satisfactory formulation of the
general principle of relativity. He had thus discarded these field
equations, being physically ill-posed (see later). Importantly,
these are also logically unacceptable \cite{ppt}.

Certain cosmological solutions \cite{cosmo} of Einstein's %field
equations imply an ``expanding'' space. In view of the logical
unacceptability of these equations, it is then not surprising that
the result of the Braxmaier et al.\ experiment \cite{braxmaier} is
inconsistent \cite{lieu-greg} with the implied expansion of the
space.

Now, the general principle of relativity ``looks'' for {\em all\/}
the laws of physics that are the {\em same\/} for {\em all\/} the
observers, undergoing whatsoever relative motions. Then, these
laws of physics will also include the laws of the quantum. This is
the reason why we can expect that a suitable theory based on this
principle of relativity fundamentally includes the phenomena also
from the quantum world.

Clearly, the general principle of relativity then seeks also for
{\em all\/} the laws about {\em all\/} the permissible
interactions of physical bodies.

The general principle of relativity is a basis \cite{unify} for
the unification of interactions then. It is a basis provided that
we develop a suitable mathematical framework for it.

In this connection, Einstein had expressed \cite{yale-oup} his
valued judgement about the general principle of relativity that
the idea of general relativity ``is a purely formal point of view
and not a definite hypothesis about nature ...''.

What he means here is that the general principle of relativity is
not a statement about the (character of the laws of) Nature, but
it is rather that this is a statement about how one should, as a
strategy, obtain the laws of Nature.

Then, rather than restricting ourselves to the confines of the
special principle of relativity, it is ``advisable'' to formulate
a theory on the basis of the general principle of relativity. This
should get viewed indeed as a statement about a definite strategy
on as to how to obtain the laws of Nature, and not as a hypothesis
about Nature.

Within this strategy of the general principle of relativity, it is
then not pivotal as to how many species of (elementary) particles
exist; what their specific properties as well as
interrelationships are, how many fundamental forces Nature has,
etc.

As long as we have a satisfactory mathematical framework for the
general principle of relativity, these above (and results alike)
could then be looked upon as some ``predictions'' of the
corresponding theoretical framework. Essentially, a theory (as a
conceptual and the corresponding mathematical framework) will have
to tell us what its ``observables'' are.

Moreover, such predictions of a ``theory of everything'' will be
``verifiable'' and could then establish or demolish the chosen
mathematical premise of the general principle of relativity, but
not the principle, which is a purely formal point of view. It
therefore follows that if one mathematical formalism fails to
explain observations, then we replace it by suitable other.

This is the meaning of the general principle of relativity being a
purely formal point of view then. This is also why Einstein had
been searching for a suitable mathematical framework to represent
this principle, and why he had expressed some doubts \cite{pais}
[p. 8] as to whether the standard differential geometry would help
with further progress. (See later.)

It is well known \cite{schlipp, pais} that Einstein attempted
numerous formulations for what he called the unified field theory.
All these formulations were however unsatisfactory even for their
creator. These later years of his scientific career often get
described \cite{pais} [p. 341] as: ``he was as clear about his
aims as he was in the dark about the methods by which to achieve
them.''

To understand the problem that so persistently troubled Einstein,
let us represent a material body as a point of the 4-dimensional
spacetime geometry. This we do because Newton's theory so
represents a physical body, explains some phenomena and we demand
that our field equations reduce to the appropriate equations of
Newton's theory in some suitable approximation. Moreover, it is
also imperative that any theory of everything resolves this
problem satisfactorily.

Now, let a point that is a curvature singularity of the geometry
represent a physical body as in, {\it eg}, Schwarzschild's famous
solution of Einstein's equations. Any mathematical expressions
valid at other points of geometry do not hold at its curvature
singularity. Then, the ``motion'' of a curvature singularity is a
``singular'' curve and it is not part of that geometry. Therefore,
equations for a curve of the geometry, geodesic equations, do not
provide a law of motion for the singularity. In the absence of any
law of motion for point objects, it is not possible to define
their fluxes and, hence, the energy-momentum tensor. In this case,
Einstein's equations do not have any physical meaning whatsoever.
Einstein had recognized this problem of the ``pure gravitational
fields'' as the problem of locations in space where his field
equations of general relativity are not valid \cite{schlipp,
pais}.

Next, let a chosen point of geometry be not its any curvature
singularity. Now, geometric equations (geodesic, geodesic
deviation) acquire {\em physical meaning\/} only when any point of
geometry at which they are satisfied is ascribed mass, charge etc.
This is indeed the same as Newton's theory using Euclidean
geometry and ascribing mass, charge to a point on the curve of
that geometry with equations of the curve providing Newton's laws
of motion.  In the case of Einstein's equations, only the geometry
is non-Euclidean. This is then the only relevant difference in the
two cases.

Now, an unambiguous way to ascribe mass etc.\ to any point of the
geometry is needed to define the energy-momentum tensor. In
Newton's theory (Special Theory of Relativity), this prescription
is ``by hand'' for any point of the Euclidean (Minkowskian)
geometry. How then for Einstein's case of a curved 4-dimensional
geometry is this unambiguously done?

Following Newton, we may prescribe, by hand, values of mass,
charge etc.\ for any non-singular point on the curve of a (curved)
four-dimensional geometry, formally define the energy-momentum
tensor, write Einstein's field equations and obtain their
(spacetime) solutions.

Where is the problem then? The problem is that the mass of a
physical body is represented by the curvature of the spacetime
geometry. ``More the matter, the larger is the curvature.'' This
leads \cite{mtw} to the problem of defining the energy (or mass)
of a system in general relativity. As is well known \cite{mtw},
there is no satisfactory definition of mass in general relativity
and, hence, the above prescription of mass for a point of curved
geometry cannot be made in any unambiguous manner. Hence, mass
cannot be the part of the energy-momentum tensor in the above
simple way.

By comparison to this trouble of Einstein's field equations,
Newton's theory faces no such difficulty. This is so because mass,
charge etc.\ are the ``extraneous'' mathematical quantities, which
are independent of the nature of the Euclidean geometry underlying
Newton's theory. These characteristics can, therefore, be
specified, by hand and at will, for a point of that geometry.

But, unambiguous prescription of mass, charge etc.\ for a point of
the curved geometry becomes {\em impossible\/} when we attribute
the strength of the gravitational field to the curvature of the
(spacetime) geometry. Einstein expressed \cite{schlipp} this
difficulty in the words that the ``right side'' of his equations
is ``a formal condensation of all things whose comprehension in
the sense of a field theory is still problematic.'' That is to
say, his field equations are physically ill-posed.

We also note that a curved (spacetime) geometry is not the unique
mathematical notion implied by Einstein's equivalence principle,
or by the general principle of relativity; the latter being the
actual basis of general relativity. In fact, the general principle
of relativity does not ``prescribe'' any specific mathematical
framework for itself.

This ``explains'' why Einstein explored \cite{pais} the question
of whether the fundamental mathematical basis for physics might be
other than that of the partial differential equations and had
expressed \cite{pais} [p. 8] doubts as to whether differential
geometry be used for further progress \footnote{Within the
mathematical framework of Standard Differential Geometry,
``values'' of mass, charge etc.\ can only get ``specified by
hand'' for a point of the geometry. Mathematical methods away from
differential geometry are, therefore, needed for appropriately
``representing origins'' of the physical conceptions of inertia
etc.}.

The pivotal issue here is then of an appropriate mathematical
framework for the ideas behind the general principle of
relativity. The following then appears to be a relevant step in
the direction of this mathematical framework.

Historically, Descartes \cite{ein-pop} was the first to express
concern that a coordinate system of the Euclidean space of the
Newtonian Mechanics, as a ``material construction'' of a reference
body, does not get affected by physical precesses. Newton
recognized this in the form of the ``absolute space'' underlying
his Mechanics. Systems of reference in Newton's theory and in
special relativity, both, cannot be any material bodies then. This
is a definite drawback of these basic theories.

Unless we address Descartes's above issue, no mathematical
framework for the general relativity would be satisfactory. Hence,
the {\em general\/} principle of relativity must deal with the
``physical'' reference systems, physical bodies, undergoing
whatever relative motions. This principle of relativity can
therefore find its proper mathematical representation only when we
have an appropriate mathematical structure to represent all the
physical bodies usable as reference systems.

Changes in the physical bodies are the {\em physical phenomena}.
As only physical bodies get used as reference in physical
measurements, mathematical structure(s) representing physical
bodies will have to be such that phenomena become `changes' to
(transformations  \footnote{Dirac [(1939) {\it Proc R Soc
(Edinb)}, {\bf 59}, 122] had emphasized the importance of
transformations as: ``... transformations play an important role
in modern physics; both relativity and quantum theory seeming to
show that transformations are of more fundamental importance than
equations.''} of mathematical structure(s) of) reference systems
themselves.

Then, a mathematical framework, which is different than that of
the standard differential geometry, that properly represents
physical bodies and that is therefore consistent with the general
principle of relativity is needed \cite{utr} for further progress.
This mathematical framework needs to be based on the concept of a
Quasi-Category \cite{unify} and, as such, was called the Universal
Relativity to differentiate it from Einstein's general relativity,
both being based on the general principle of relativity.

As discussed here, the result of the Braxmaier et al experiment
\cite{braxmaier} appears to broadly support (certain parts of)
Einstein's viewpoints of later years \cite{schlipp, pais} and also
the general approach together with the mathematical methodology in
\cite{unify, utr}, both. This approach that utilizes the category
theoretical methods to represent the general principle of
relativity is then a promising program for the unification of all
the fundamental physical interactions.

We therefore conclude that the Braxmaier et al experiment
\cite{braxmaier} rules out Einstein's field equations but supports
the general principle of relativity, a strategy for developing a
theory.

%%%%%%%%%%%%%%%%%%%%%%%%%%%%%%%%%%%%%%%%%%%%%%%%%%%%%%%%%%%%%
%\newpage
%\bigskip

\end{document}